\pdfoutput=1
\documentclass[apj]{emulateapj}
\usepackage{apjfonts}
\usepackage[bookmarks=true,colorlinks=true,citecolor=blue,linkcolor=magenta,urlcolor=cyan]{hyperref}

\usepackage{natbib}
\usepackage{amsmath}
\usepackage{amssymb}
\usepackage{subfigure}
\usepackage{url}
\usepackage{CJK}

\renewcommand{\vec}[1]{ {\mathbf #1} }

\newcommand{\curl}{ {\bf \nabla} \times}

\newcommand{\jcphy}{  {\it J.~Comput.~Phys.}}

\newcommand{\Eq}{{Equation}}

\newcommand{\Fig}{{Figure}}
\newcommand{\Figs}{{Figures}}

\newcommand{\divB}{\nabla\cdot\mathbf{B}}
\newcommand{\crlB}{\nabla\times\mathbf{B}}

\shorttitle{Comparison of Coronal Magnetic Field Models}
\shortauthors{Duan et al.}

\begin{document}
\begin{CJK*}{UTF8}{gbsn}

\title{Comparison of Two Coronal Magnetic Field Models for
    Reconstructing a Sigmoidal Solar Active Region With Coronal Loops}

\author{
  Aiying Duan \altaffilmark{1,2,4},
  Chaowei Jiang \altaffilmark{3,2,4},
  Qiang Hu \altaffilmark{4,5},
  Huai Zhang \altaffilmark{1},
  G. Allen Gary \altaffilmark{4},
  S.~T. Wu \altaffilmark{4},
  Jinbin Cao \altaffilmark{6}}

\altaffiltext{1}{Key Laboratory of Computational Geodynamics, University
  of Chinese Academy of Sciences, Beijing 100049, China}

\altaffiltext{2}{SIGMA Weather Group, State Key Laboratory for Space
  Weather, National Space Science Center, Chinese
  Academy of Sciences, Beijing 100190, China}

\altaffiltext{3}{HIT Institute of Space Science and Applied Technology, Shenzhen, 518055, China}

\altaffiltext{4}{Center for Space Plasma and Aeronomic Research, The
  University of Alabama in Huntsville, Huntsville, AL 35899, USA}

\altaffiltext{5}{Department of Space Science, The University of Alabama in Huntsville, Huntsville, AL 35899, USA}

\altaffiltext{6}{School of Space and Environment, Beihang University, Beijing 100191, China}

\email{duanaiying@ucas.ac.cn, hzhang@ucas.ac.cn, chaowei@hit.edu.cn}

\begin{abstract}
  Magnetic field extrapolation is an important tool to study the
  three-dimensional (3D) solar coronal magnetic field which is
  difficult to directly measure. Various analytic models and numerical codes
  exist but their results often drastically differ. Thus a critical comparison of
  the modeled magnetic field lines with the observed coronal loops is
  strongly required to establish the credibility of the model. Here we compare two different non-potential extrapolation codes, a non-linear force-free field code
  (CESE--MHD--NLFFF) and a non-force-free field (NFFF) code in
  modeling a solar active region (AR) that has a sigmoidal
  configuration just before a major flare erupted from the region. A
  2D coronal-loop tracing and fitting method
  is employed to study the 3D misalignment angles between the
  extrapolated magnetic field lines and the EUV loops as imaged by
  SDO/AIA. It is found that the CESE--MHD--NLFFF code with preprocessed
  magnetogram performs the best, outputting a field which
  matches the coronal loops in the AR core imaged in AIA~94~{\AA} with
  a misalignment angle of $\sim 10^{\circ}$. This suggests that
  the CESE--MHD--NLFFF code, even without using the information of coronal loops in
   constraining the magnetic field, performs as good as some coronal-loop forward-fitting models.
  For the loops as imaged by AIA~171~{\AA} in the outskirts of
  the AR, all the codes including the potential-field give comparable results of mean misalignment angle ($\sim~30^{\circ}$).
  Thus further improvement
  of the codes is needed for a better reconstruction of the long loops enveloping the
  core region.
\end{abstract}


\section{Introduction}
\label{sec:intro}

The observed solar explosive events (e.g., flares and coronal mass ejections) within the corona is attributed to  an energy release in the coronal magnetic field. Hence understanding the dynamics and physics of these eruptions is centered on understanding the structure and evolution of the magnetic field. The coronal field can evolve from reconnection and photospheric variations. However, direct measurement of the coronal magnetic field from
emissions of the extremely tenuous plasma of the corona by
spectropolarimetric methods still proves to be
difficult~\citep{LinH2004, LinH2016}.
Currently, the routine measurement of solar magnetic field
that we can rely on is restricted to only a single layer of solar surface, i.e., the photosphere.
Most recently, chromospheric polarimetry is beginning to show promising results, however the height of formation is problematic~\citep{Noda2017}.

Due to the lack of measurement data, the three-dimensional magnetic
field in the solar corona is usually ``extrapolated'' or
``reconstructed'' in numerical ways from the photosphere surface data
based on particular assumptions or models. Such techniques of modeling
the coronal magnetic field have been developed~\citep[e.g., see review papers of][]{Sakurai1989,
  McClymont1997, Solanki2006, Wiegelmann2008, Wiegelmann2012solar,
  Regnier2013}. Three magnetic field models often used
include the potential field, linear force-free field (LFFF), and
nonlinear force-free field (NLFFF). All these models are derived from a basic
assumption that the Lorentz force in the corona vanishes in
the case of extremely low plasma $\beta$ ($\beta$ is the ratio of the plasma pressure to the magnetic pressure) and quasi-static equilibrium
of the coronal field. Consequently the electric current $\vec J =
\crlB$ must be parallel to the magnetic field, i.e., $\vec J =
\alpha \vec B$ where $\alpha$ is called the force-free parameter. In the
potential field model, $\alpha=0$; in the LFFF, $\alpha$ is a constant;
and in the NLFFF, $\alpha$ is variable in space. The earliest models were
based on potential field~\citep{Altschuler1969, Sakurai1982} and
   LFFF~\citep{Seehafer1978} that are extrapolated from only the data of
line-of-sight (LoS) component of the photospheric field since
the transverse components were not measured. Now both the
LoS and transverse components of the photospheric magnetic field can be measured ~\citep[e.g.,][]{Hoeksema2014} and information of the electric current passing through the photosphere can derived.
Nonlinear force-free field (NLFFF) reconstructions, which are based on the vector
magnetograms, are more robust than those earlier models, and employ several numerical schemes~\citep{Sakurai1981,Yang1986,Wu1990, Amari1997, Yan2000,
  He2006, Wiegelmann2006, Wheatland2006,Valori2010,Jiang2012apj}. Existing NLFFF codes include the optimization
method~\citep{Wheatland2000, Wiegelmann2004, Wiegelmann2006,
  Wiegelmann2008}, the magneto-frictional method~\citep{Valori2007,
  Valori2010, Guo2016a}, the Grad-Rubin method~\citep{Amari2006, Wheatland2007,
  Amari2014nat}, and the MHD-relaxation code based on
conservation-element/solution-element space-time
scheme~\citep[CESE-MHD-NLFFF:][]{Jiang2011, Jiang2012apj,
  Jiang2013NLFFF}. Now NLFFF models are widely
used in the solar physics community for exploring the 3D coronal
structure prior to and post solar eruptions and for understanding the
influence of the magnetic topology on solar
eruptions~\citep[e.g.,][]{Guo2008, Sun2012, Jiang2013MHD, ChengX2014,
  LiuC2014, XueZ2016}.

An alternative to NLFFF extrapolation is an approach based on a
magnetic field model of the corona derived from the variational
principle of minimum energy dissipation rate~\citep{HuQ2008, HuQ2008ApJ, HuQ2010JASTP}. The formula governing
the coronal magnetic field is more complex, and the solution,
expressed as the superposition of three LFFFs (one
being potential), is in general not force-free. Such a non-force-free
field (NFFF) extrapolation was presented in~\citet{HuQ2008} and was
shown to be applicable to NLFFF configuration given by
\citet{Low1990} analytic solutions. It was also tested on numerical
MHD simulation data for an active region~\citep{HuQ2008ApJ}. Later it
was further developed for practical applications to photospheric
vector magnetograms obtained within an active
region~\citep{HuQ2010JASTP}.  The implementation of the algorithm is
relatively simple and the full code written in Interactive Data Language (IDL) can handle the  magnetograms of 1024$\times$1024 pixels on a desktop
PC. However extensive testing and application of the NFFF algorithm
have yet to be performed.

A critical assessment of the reliability of coronal magnetic-field
extrapolation modeling is to examine the goodness-of-matching of
the geometry of the simulated magnetic field lines with that of the
observed EUV coronal loops. This is because the plasma emission in the
corona reflects the geometry of the invisible magnetic field, as in
most parts of the corona the plasma is ``frozen'' with the magnetic
fields.  While earlier studies compared theoretical models with
observed images in a rather qualitative way, recent studies go into
quantitative measurements of the agreement between modeled field
lines and observed coronal loop geometries.
In an assessment of a variety of popular NLFFF codes for modeling
AR~10953, \citet{DeRosa2009} compared the model field lines to
3D trajectories of coronal loops, which are observed
stereoscopically by the twin Solar Terrestrial Relations Observatory
(STEREO) spacecraft~\citep{Aschwanden2008}. It was found that the
misalignment angle between the field lines and the loops amounts to
rather large values of $24^{\circ}$--$44^{\circ}$, and unexpectedly,
none of the examined NLFFF models improved significantly upon the value found for the potential field model. Another method employs a slight variation by measuring the distance between a loop and a projected field line~\citep{ChaeJ2005, LimE2007, Malanushenko2011}.
 The same idea of a ``distance'' has also been used in 3D, with  STEREO data~\citep{Wiegelmann2002}.
Without tracing the trajectories of coronal loops, \citet{Wiegelmann2012} compared the model field lines with SDO/AIA images by calculating the
sum of gradient of the AIA image intensity along each projected
field lines, and the result is assumed to reach its minimum if all the
field lines are co-aligned with the corresponding loops.
The LFFF parameter $\alpha$, derived from visual fit to loops has been compared to that derived from a vector magnetogram~\citep{Burnette2004} providing an additional goodness metric.

Recently, \citet{Gary2014} developed a new method of deriving the 3D
structure of observed 2D coronal loops independent of
heliostereoscopy, and suggested that it can determine the
matching of extrapolated magnetic fields with coronal loops. In
that method, an automated loop recognition scheme
\citep[OCCULT-2,][]{Aschwanden2013OCCULT} is first used to extract
2D loop structures from EUV images and then the extracted loops are
fitted with 2D cubic B{\'e}zier splines that are based on 4 control
points~\citep{Gary2014a}. The 2D splines are further extended to 3D
with the heights of all 4 control points set as free parameters. The heights are determined by minimizing the misalignment angles of the 3D splines
with field lines of a given model of coronal magnetic field, and the resulting 3D
splines are regarded as the trajectories of the corresponding coronal
loops in 3D. So the image of these loops represents a 2D projection of these 3D structures.
Naturally, different magnetic field models will result in
different sets of misalignment angles, and a perfect matching
model can presumably yield a misalignment angle of zero. Thus this
method provides a powerful tool of comparing different magnetic-field
models of the corona by using EUV images obtained from one single viewpoint.

Other coronal magnetic field models with
forward fitting of coronal loops has been developed
for better reproducing the geometry of the loops
~\citep{Malanushenko2012, Malanushenko2014, Aschwanden2013NLFFF1, Aschwanden2013ApJ}. For instance, unlike the traditional NLFFF codes that
extrapolate the coronal field from the vector field at the
photosphere, the coronal-loop forward-fitting quasi-NLFFF models of \citet{Aschwanden2013NLFFF1} use only the LoS magnetic
field component and the non-potentiality of the field (characterized
by the force-free parameter $\alpha$) is determined by minimization
of misalignment of the field lines with the traced loops from observed
images. \citet{Aschwanden2013ApJ} shows that their code can be applied
to either 3D loops determined stereoscopically or simply 2D from LoS
observations. In the latter case, the 3D height of the loops is used
as a free parameter and a 2D misalignment angle is minimized. The
forward-fitting method strongly depends on the recognition of coronal
loops from the image, which has been automated by
\citet{Aschwanden2013OCCULT}. We should note that even in such
forward-fitting of the coronal loops, the misalignment angles of the field
line with the loops are significant. For example, the
misalignment angle resulted in \citet{Aschwanden2013ApJ} is about
$20^{\circ}$ in the forward-fitting of TRACE loops.

In this paper, we will evaluate our extrapolation results from the
NLFFF (CESE--MHD--NLFFF) and the NFFF codes, by introducing a critical
comparison of the modelled magnetic field lines with the coronal loop
geometry. In particular, we will
employ \citet{Gary2014}'s method to compute the misalignment angles of
the model field lines and identified loops. The target region used in
this examination is AR~12158 prior to its major eruption on 2014
September 10, which exhibits a well-shaped sigmoid that indicates a
non-potential field. Employing the SDO AIA 94 Angstroms images, we find that one NLFFF solution fits the best
(misalignment angle of $\sim 10^{\circ}$) the AR core loops, but not so well the loops that correspond to large overlying
field lines (misalignment angle of $\sim 30^{\circ}$) near the side boundaries.
Our results are similar to those of \citet{DeRosa2009}.  In addition
to the comparison of field lines with coronal loops, we also evaluate
quantitatively the extrapolated fields by other means including the
magnetic energy, helicity contents, magnetic twist and squashing degree~\citep{Demoulin1996, Titov2002}.  The paper is organized as follows. We first
briefly describe the CESE--MHD--NLFFF code and the NFFF code in
Sections~\ref{sec:model} and \ref{sec:model1}, respectively. Then in
Section~\ref{sec:loop} we give a short review of the coronal loop
tracing and fitting method developed by \citet{Gary2014}. The results
of assessing the extrapolated coronal magnetic fields for AR~12158 are
given in Section~\ref{sec:res}, and finally discussions appear in
Section~\ref{sec:con}.

\section{The CESE--MHD--NLFFF code}
\label{sec:model}

The CESE--MHD--NLFFF code~\citep{Jiang2013NLFFF} belongs to the class
of MHD relaxation methods. It solves a set of modified zero-$\beta$ MHD
equations with frictional force using an advanced
conservation-element/solution-element (CESE) spacetime scheme on an
non-uniform grid with parallel computing~\citep{Jiang2010}. The
modified MHD equations are written as
\begin{eqnarray}
  \label{eq:main}
  \frac{\partial\rho\mathbf{v}}{\partial t} =
  (\crlB)\times\mathbf{B}-\nu\rho\mathbf{v}, \ \
  \rho=|\vec B|^{2},\ \
  \frac{\partial\mathbf{B}}{\partial t} =
  \nabla\times(\mathbf{v}\times\mathbf{B})
\end{eqnarray}
where $\nu$ is the frictional coefficient. The initial condition for
the computation sequence is a potential field extrapolated from the vertical
component of the vector magnetogram.
To drive the evolution of the field, we change
the horizontal magnetic components at the bottom boundary gradually
until they match the vector magnetogram, after which the system will
be relaxed to a new equilibrium. Details of this code can be found in~\citep{Jiang2012apj, Jiang2012apj1}. It is well
tested by different benchmarks including the \citet{Low1990}'s
analytic force-free solutions and the \citet{Titov1999}'s magnetic
flux rope model, and recently applied to the SDO/HMI vector
magnetograms~\citep{Jiang2013NLFFF, Jiang2014NLFFF}.  Here we use the
code in exactly the same way as~\citet{Jiang2013NLFFF}, without any
parameter optimization for the present modeling.

Unlike the coronal field, the photospheric field is not necessarily
force-free because of much higher plasma $\beta$, thus it is usually
required to remove the Lorentz force in the vector magnetogram for a
boundary field consistent with the force-free assumption of
NLFFF extrapolations~\citep{Wiegelmann2006}. Meanwhile, smoothing of original magnetogram
is needed to reduce the data noise and smooth the very small-scale structures that
cannot be properly resolved by the discretized grid. Such process of removing force and smoothing
is called preprocessing
and here we use the preprocessing code
developed by~\citet{Jiang2014Prep}.  Different from other
preprocessing codes~\citep{Wiegelmann2006, Fuhrmann2007,
  Fuhrmann2011}, this code is unique in that it splits the vector
magnetogram into a potential field part and a non-potential field part
and handles the two parts separately. The potential
part, as it is already force-free, only needs smoothing, which is simply performed
by taking the data sliced at a plane one
pixel (of an HMI magnetogram) above the photosphere from the 3D
extrapolated potential field. Then the non-potential part is modified and smoothed by an
optimization method to fulfill the constraints of total magnetic
force-freeness and torque-freeness. One advantage of using such
a splitting is that the preprocessing of the non-potential field
part can be guided by the extents of force-freeness and smoothness of
the smoothed potential-field part. This is because in the
practical computation, particular attention needs to be paid on
what extent the force needs to be removed and the smoothing can
be performed. In practical computation based on numerical
discretization, an accurate satisfaction of force-free constraints is
apparently not necessary. Also the extent of the smoothing for the
data needs to be carefully determined, if we want to mimic the
expansion of the magnetic field from the photosphere to some specific
heights. Over-smoothing of the data may smear the basic
structures while insufficient smoothing cannot filter the small-scale
noise sufficiently. A careful choice of the weighting factors $\mu$ is
required to deal with these problems. In \citet{Jiang2014Prep} the
values of force-freeness and smoothness calculated from the
preprocessed potential-field part are used as a reference, and the
target magnetogram is required to have the same level of force-freeness
and smoothness as its potential part in numerical precision. It is found that these
requirements can restrict well the free parameters, i.e., the weighting
factors $\mu$ in the optimization function.

\section{The Non-Force-Free Extrapolation code}
\label{sec:model1}

The non-force-free field is governed by the following
equation~\citep{HuQ2008,HuQ2010JASTP}:
\begin{equation}
  \nabla\times\nabla\times\nabla\times {\bf B}+
  a \nabla\times\nabla\times {\bf B}+ b\nabla\times{\bf B}=0.
\label{nffmag}
\end{equation}
One solution is written
$\mathbf{B}=\mathbf{B}_1+\mathbf{B}_2+\mathbf{B}_3$, where each
sub-field $\mathbf{B}_i$ satisfies the standard LFFF equation with distinct parameters $\alpha_i, i=1,2,3,$
such that $a=-(\alpha_1+\alpha_3)$ and $b=\alpha_1\alpha_3$. To obtain
a solution of this form, taking advantage of the relatively simple
solutions to LFFFs, it turns out that $\alpha_2=0$, and
$\alpha_1\ne\alpha_3\ne 0$. Therefore, one of the sub-fields,
$\mathbf{B}_2$, becomes potential field solution for the boundary conditions.

For completeness, we give below a brief description of the algorithm
of deriving an NFFF solution to equation~(\ref{nffmag}) via the
superposition of three LFFFs of distinct $\alpha$ parameters. It can
be shown that the result is the following equation,
\begin{equation}
  \left( \begin{array}{c} \mathbf{B}_1 \\
      \mathbf{B}_2 \\
      \mathbf{B}_3 \end{array} \right)=
  \mathcal{V}^{-1}\left( \begin{array}{c} \mathbf{B} \\
      \nabla\times\mathbf{B} \\
      \nabla\times\nabla\times\mathbf{B}\end{array} \right) .
\label{bvc}
\end{equation}
Here the matrix $\mathcal{V}$ is composed of elements $\alpha_j^{i-1},
i,j=1,2,3$, which is a Vandermonde matrix and is guaranteed invertible
as long as the $\alpha$'s are distinct~\citep{HuQ2008}. Therefore each LFF sub-field
can be solved for the known $\alpha$ parameter, and the given normal
boundary condition by using a standard
LFFF solver~\citep{Alissandrakis1981}. Ideally the bottom boundary
condition, as given by the right-hand side of \Eq~(\ref{bvc}), has
to be derived by utilizing two or more than two layers of vector
magnetograms since the vertical gradient as well as the transverse
gradients of magnetic field have to be calculated.
If multiple layer vector magnetograms are available, then the right-hand side of \Eq~(\ref{bvc}) can provide the boundary conditions (vertical components) for each sub-field, given known $\alpha$ parameters. So the optimal pair of $(\alpha_1,\alpha_3)$ parameters, while keeping $\alpha_2\equiv 0$, is determined by a trial-and-error process by finding a pair which minimizes the average deviation between the observed ($\mathbf{B}_t$)  and the calculated ($\mathbf{b}_t$) transverse field, as indicated by the following metric~\citep{HuQ2008}:
\begin{equation}
  E_n=\sum_{i=1}^M|\mathbf{B}_{t,i}-
  \mathbf{b}_{t,i}|/\sum_{i=1}^M|\mathbf{B}_{t,i}|.
  \label{En}
\end{equation}
However, since most vector magnetograms are only available on the photosphere, which only allows for an evaluation of $(\nabla\times\mathbf{B})_z$, the boundary conditions for each sub-field cannot be provided by the third-order system of \Eq~(\ref{bvc}). An algorithm was devised by~
\citet{HuQ2010JASTP} to work with the single layer vector magnetogram by adding an additional round of iteration over successive correction to the potential sub-field $\mathbf{B}_2$. Starting with an initial guess, e.g., the simplest being $\mathbf{B}_2=0$, the system of \Eq~(\ref{bvc}) is reduced to 2nd-order which allows for the determination of boundary conditions for $\mathbf{B}_1$ and $\mathbf{B}_3$, and thus the normal trial-and-error process as described above. If the resulting minimum $E_n$ value is not satisfactory, then a corrector potential field to $\mathbf{B}_2$ is derived from the difference transverse field, i.e., $\mathbf{B}_t-\mathbf{b}_t$, and added to the previous $\mathbf{B}_2$, in anticipation of improved match between the transverse fields, as measured by $E_n$. The positive effect of such successive corrections is generally demonstrated by a monotonic decrease in $E_n$ from an initial value around 1.0 to less than 0.3, after 10,000 steps~\citep{HuQ2010JASTP}.

The algorithm relies on the implementation of fast calculations of the
LFFFs including the potential field, such as the classic algorithm of
Alessandrakis via Fast Fourier Transform (FFT) for an active region in
a Cartesian box with periodic boundary conditions. It is highly desirable to extend the scheme to the whole
solar sphere, by taking advantage of the recently developed fast
algorithm of~\citet{Jiang2012SoPh} for global LFFF (including
potential) extrapolation based on FFT as well. This extension also overcomes the limitation of flux imbalance intrinsic to the LFFF within a finite AR. This would contribute to the errors in our current NFFF extrapolation of AR magnetic field, which is based on LFFF solutions.

\section{Coronal loop tracing and fitting}
\label{sec:loop}

Here we briefly review the procedures of tracing and fitting the
coronal loops developed by~\citet{Gary2014}. First, an automated loop
recognition scheme (the code OCCULT-2) is employed to trace the loop
structures from the EUV images taken by SDO/AIA in 171~{\AA}. The OCCULT-2 code was
developed by \citet{Aschwanden2013OCCULT} and is included in the
SolarSoftWare (SSW), named as ``looptracing\_auto4.pro'', and here we
use the version dated 8-Dec-2015. The code provides a group of input
parameters to control the tracing, which are given here respectively
following~\citet{Aschwanden2014}: the low-pass-filter constant $n_{\rm
  sm1}=5$, the minimum loop curvature radius $r_{\rm min}=30$, the
minimum loop length $l_{\rm min}=20$, $n_{\rm loop}=50$ and the base
level factor $q_{\rm med}=5$.  The code outputs a number of 2D curves
corresponding to the 2D loop identified from the EUV images. Then each of
the traced 2D loops is fitted by a 2D cubic B{\'e}zier spline, which
is defined by four control points as~\citep{Gary2014a,Gary2014}
\begin{eqnarray}
  \label{eq:3D-bezier}
  \vec R(u) = [(1-u)^{3}x_{1}+3u(1-u)^{2}x_{2}+3u^{2}(1-u)x_{3}+u^{3}x_{4},
  \nonumber \\
  (1-u)^{3}y_{1}+3u(1-u)^{2}y_{2}+3u^{2}(1-u)y_{3}+u^{3}y_{4},
  \nonumber \\
  (1-u)^{3}z_{1}+3u(1-u)^{2}z_{2}+3u^{2}(1-u)z_{3}+u^{3}z_{4}]
\end{eqnarray}
where $u \in [0,1]$ is a parameter along the curve,
$(x_{j},y_{j},z_{j})$ with $j=1,2,3,4$ are the coordinates of the four
control points, and here $z_{j}=0$ and so $z=0$ as for a 2D curve. The fitting is
realized by minimizing the RMS distances of 10 equally-spaced points
along the loop curve and the B{\'e}zier spline.
It was proven that generally there is no need to
use a higher-order curve than cubic B{\'e}zier curve (of 3rd order),
which can sufficiently fit the EUV loops~\citep{Gary2014a,
  Gary2014}. After this, the 2D B{\'e}zier splines are extended to 3D
by determining the four non-zero $z_j$ coordinates for which the curve
(\Eq~\ref{eq:3D-bezier}) has minimal misalignment angle with the field vector along it.
In particular, at a
given point $\vec R$ of a B{\'e}zier curve, the misalignment angle
$\mu (\vec R)$
between the direction of the 3D spline (or the 3D coronal loop)
$\vec L (\vec R)$ and the magnetic field model $\vec B (\vec R)$ at the same point
is defined as
\begin{equation}
  \label{eq:misalign1}
  \mu(\vec R) = \arccos\left[\frac{\vec B (\vec R) \cdot \vec
      L(\vec R)}{|\vec B (\vec R)||\vec L (\vec R)|}\right],
\end{equation}
Thus for a single loop, a mean misalignment angle can be defined by
\begin{equation}
  \label{eq:misalign2}
  \xi = \frac{1}{\Gamma}\sum_{k=1}^{\Gamma}\mu(\vec R_{k}),
\end{equation}
where $k$ is node index of the curve, and here we use an evenly-spaced
100 points ($\Gamma=100$) along the loop length. As $\xi$ is a
function of the four altitudes $z_{j}$, the best-fit altitudes
$z_{j}^{*}$ are determined by minimizing $\xi$.
After applying the minimization program to all the
loops, statistical assessments of misalignment of the
model field lines and loops can then be performed.

\begin{figure*}[htbp]
  \centering
  \includegraphics[width=0.9\textwidth]{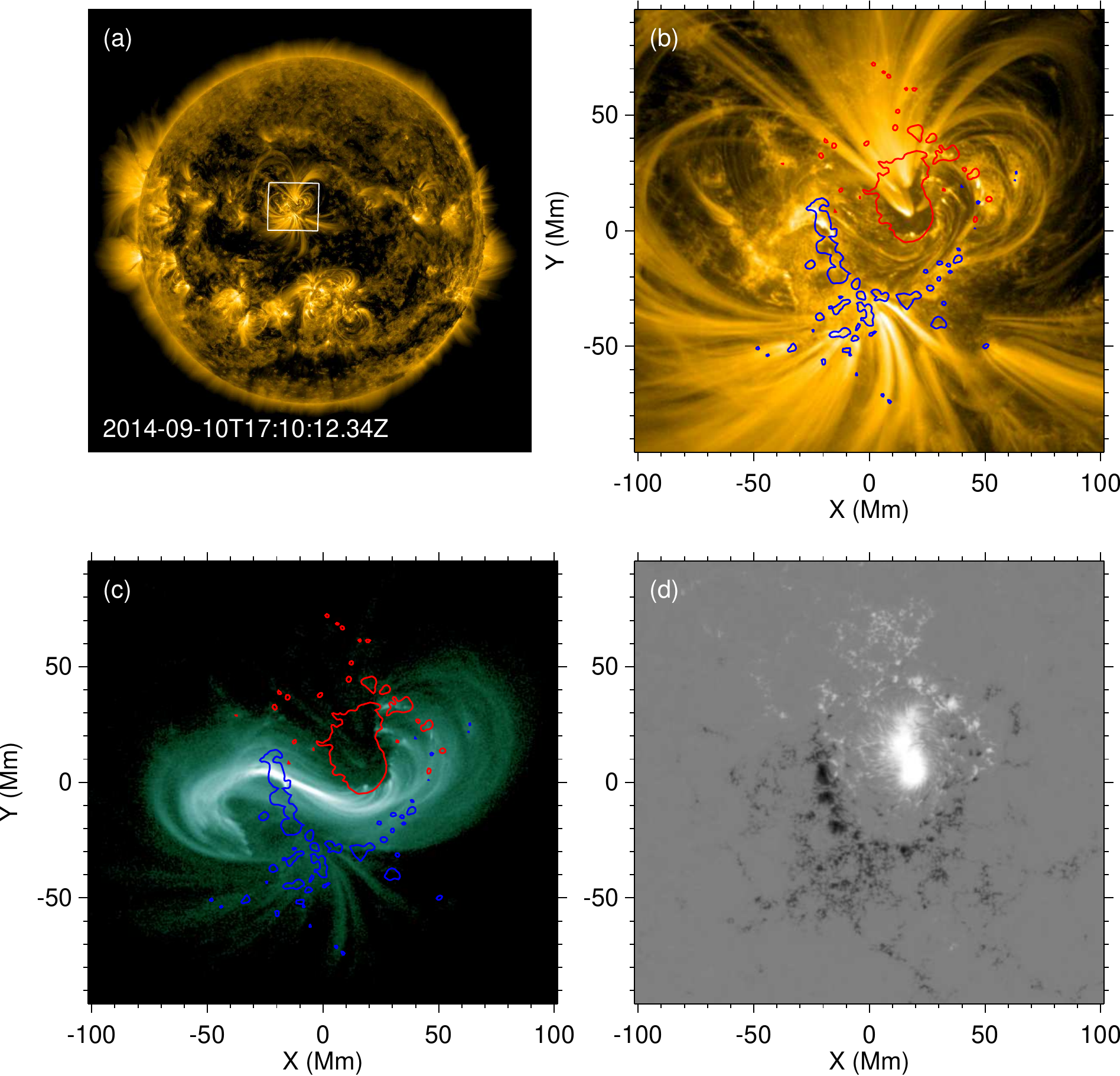}
  \caption{Observations of AR~12158 at the time 17:10~UT on 2014
    September 10. (a) AIA 171~{\AA} full-disk image with location of
    AR~12158 marked by the box. (b) Close view of the boxed
    region. (c) Same as (b) but in AIA 94~{\AA} channel. (d) SHARP CEA
    remapped LoS magnetogram.}
  \label{fig:fulldisk}
\end{figure*}

\begin{figure*}[htbp]
  \centering
  \includegraphics[width=0.9\textwidth]{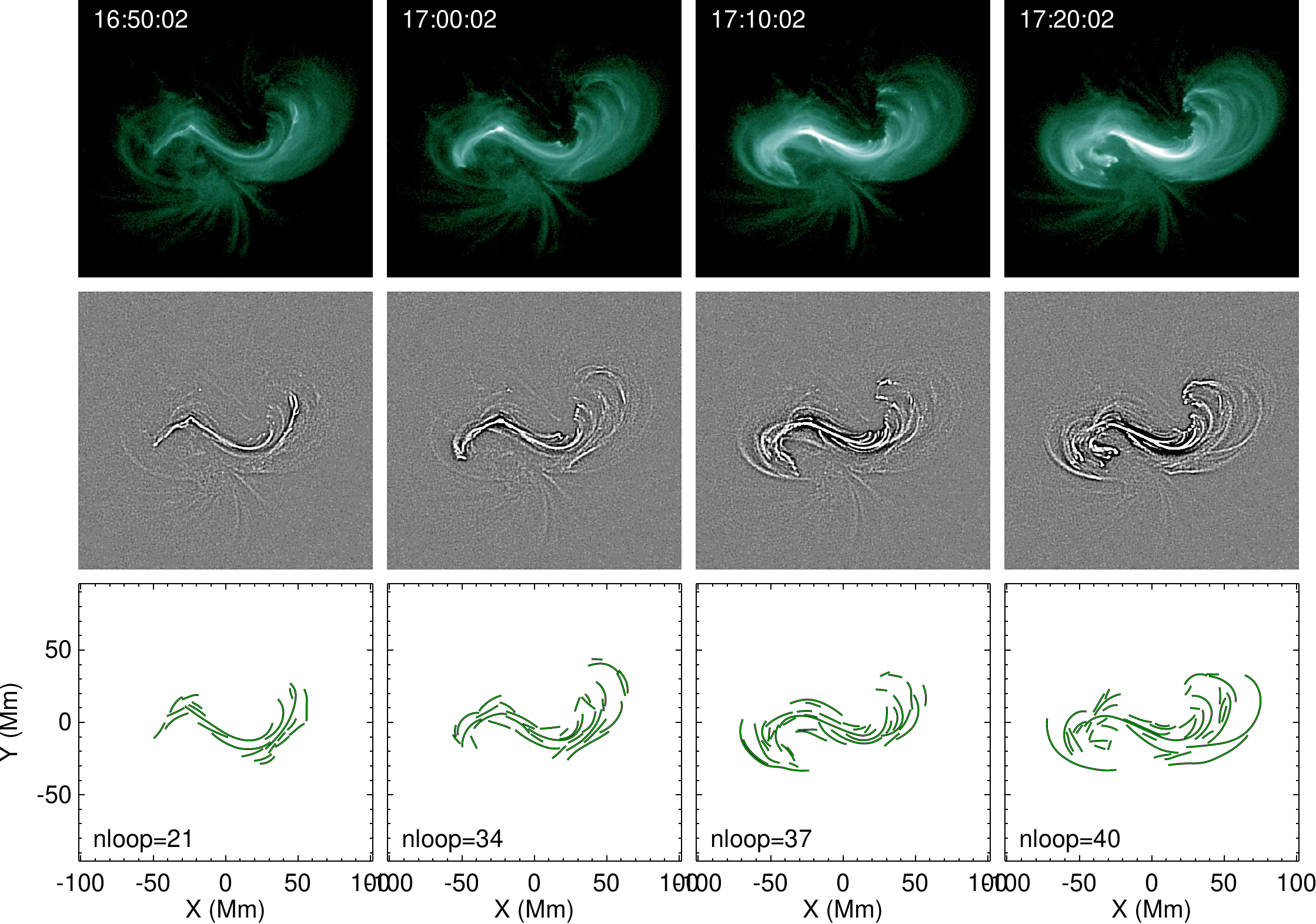}
  \caption{Top: AIA 94~{\AA} images of the sigmoid around the flare
    time. Middle: Same AIA data as the top but is bandpass filtered
    with a low-pass-filter boxcar of $n_{\rm rm sm1} = 5$ pixels and a
    high-pass-filter boxcar of $n_{\rm sm2} = 7$ pixels. Bottom:
    Automated loop tracing of the bandpass-filtered images using the
    code OCCULT-2.}
  \label{fig:selection}
\end{figure*}

\section{Results}
\label{sec:res}

AR~12158 on 2014 September 10 is selected as our target of study
because of its clear sigmoidal configuration imaged by SDO/AIA in 171~{\AA} channel as well as
its near disk-center location. \Fig~\ref{fig:fulldisk} shows that
AR~12158 was passing the central meridian on 2014 September 10, and
moreover, the region appears to be isolated from neighboring ARs,
although there are relatively long coronal loops reaching far from the
core region. An eruptive flare of X1.6 class occurred in the core of this
region, which began at 17:21~UT, reached its peak at 17:45~UT, and
ended at 18:20 UT. This sigmoid and its flare have been studied by
several authors. From observations of SDO/AIA and interface region
imaging spectrograph (IRIS), \citet{ChengX2015} suggested that prior to
the major eruption a magnetic flux rope is under formation by
tether-cutting reconnection~\citep{Moore2001} between two groups of
sheared arcades driven by the shearing and converging flows in the
photosphere near the polarity inversion line. \citet{LiT2015} reported
slipping reconnection of flaring loop during the flare process, which
was also carefully studied by \citet{Dudik2016} in the context of the
3D standard flare model~\citep{Aulanier2012, Janvier2013}. NLFFF
extrapolation of the coronal magnetic field about 2 hours before the
flare has been carried out by~\citet{ZhaoJ2016} using a Grad-Rubin
code implemented in spherical coordinates~\citep{Gilchrist2014}. By
computing a map of the magnetic field squashing factor~\citep{Titov1999},
\citet{ZhaoJ2016} suggested that the locations of QSLs
agrees to some extent with the observed flare ribbons, which are
predicted by the 3D standard flare model.
They found a strongly twisted and complex magnetic flux rope in
the extrapolation, the shape of which, however, is not seen in AIA images.

Here we perform extrapolations of the coronal magnetic
field immediately before the X-class flare onset. Specifically we use the
HMI vector magnetogram at 17:00~UT on September 10, and the
extrapolations are to be compared with the AIA observations at
17:10~UT. The reason for selecting AIA observations at 17:10~UT is
that the sigmoid-shaped emission appears to be most distinctive in the
94~{\AA} channel when we inspect the AIA images before the flare
beginning time of 17:21~UT (see \Fig~\ref{fig:selection}). As can be seen in
\Fig~\ref{fig:selection}, the loop-tracing code OCCULT-2 identifies a bundle of
loops forming an inverse S shape, which
looks most apparent at the selected time. The Space weather
HMI Active Region Patches (SHARP) vector magnetogram data product `hmi.sharp\_cea\_720s' \citep{Hoeksema2014, Bobra2014}
is used for our extrapolation. It includes vector magnetograms projected and re-mapped onto the cylindrical
equal area~\citep[CEA][]{snyder1987map} Cartesian coordinate system centred on the tracked AR,
which is well-suited for the Cartesian version of our code. For comparing the coronal loops with the
modeled field, we realign the AIA images using the same CEA remapping to
assure that the AIA images are co-aligned with the magnetogram. We note this
is reasonable only when the target region is located near the disk
center and should be small enough, such that the solar radial direction is
nearly co-aligned with the line of sight
and the 2D loops in the AIA image correspond to projection of 3D loops
along the radial direction. Otherwise,
comparison of the magnetic field lines with traced loops from EUV
images should be performed directly in spherical
geometry~\citep{Aschwanden2014} using a spherical version of the \citet{Gary2014}'s loop-fitting code, and thus extrapolation in spherical
coordinates is required~\citep[e.g.,][]{Jiang2012apj1, Tadesse2014,
  Gilchrist2014}.

For the given SHARP vector magnetogram at the selected time, we
apply three magnetic field models with different numerical method and
different boundary conditions. The first one (NLFFF1) is
CESE--MHD--NLFFF modeling using the original vector magnetogram. The
second one (NLFFF2) is CESE--MHD--NLFFF with the preprocessed
magnetogram. The third model is NFFF extrapolation, which also uses
the vector magnetogram without preprocessing as in NLFFF1.  All the
extrapolations use the same resolution of 1~arcsec, and the results
are compared within the same effective volume of $282\times
266\times 200$~arcsec$^3$.

In the following sections, we assess and compare the magnetic field
models in different ways including the magnitude of the Lorentz force
and the quality of divergence-freeness, the magnetic configuration and
topology features by computing the magnetic field lines and squashing
degree, misalignment of the magnetic field lines with the traced
loops, and finally the magnetic energy and helicity content of the
fields.

\subsection{Lorentz force and magnetic field divergence}
\label{sec:metrics}

To check the quality of force-freeness and divergence-freeness of the
reconstructed fields, two metrics are routinely
used~\citep[e.g.,][]{Schrijver2006, DeRosa2009, Jiang2013NLFFF}: the
mean sine of the angle between current $\vec J$ and $\vec B$ weighted
by $J$, named as CWsin and defined by
\begin{equation}
  {\rm CWsin} \equiv
  \frac{\int_{V}{J}\sigma dV}{\int_{V}{J} dV};
  \ \
  \sigma =  \frac{|{\vec J}\times{\vec B}|}{{J}{B}},
\end{equation}
where $B=|\vec B|$, $J=|\vec J|$ and $V$ is the computational
volume; and a normalized divergence error measured
by
\begin{equation}
  \langle |f_{i}|\rangle =
  \frac{1}{V}\int_{V}\frac{\divB}{6B/\Delta x} dV.
\end{equation}
These two metrics are equal to zero for an exact or a perfect force-free
field; hence, the smaller the metrics are, the better the
extrapolation is for a force-free field. However, for a field with very
small current, the CWsin does not necessarily give a reliable or meaningful
value because of random numerical errors. As an example, value of CWsin could
be close to $1$ for a potential field solution computed by Green's function
method or other numerical realization. The reason is that the
numerical finite difference, used for computing the current $\vec J =
\curl \vec B$ from $\vec B$, gives small but finite currents, whose
directions are randomly from 0$^{\circ}$ to 180$^{\circ}$, thus
the angle between $\vec J$ and $\vec B$ should have an average value of $\sim
90^{\circ}$. Consequently, the distribution of CWsin for a NLFFF is `contaminated' by this if a substantial portion of the volume is current-free.
The issue has been previously noted~\citep{Jiang2012apj,
  Malanushenko2014}.
Small-scale structures (mainly in the
weak field regions) in the solar magnetograms, as
are not sufficiently resolved,
also might increase the value of CWsin. This is because in these
regions, although the magnetic field strength is small, the derived
current by numerical difference might not be small and their
directions are often random. It seems to explain why usually many
NLFFF extrapolations from real magnetograms give CWsin values, for example, $\sim 0.30$~\citep{DeRosa2009}, which is much
larger than results of benchmark tests with ``idealized magnetogram''
\citep[which are $\sim 0.1$ or even smaller, see][]{Jiang2012apj}.  As
demonstrated by \citet{Jiang2013NLFFF}, the CWsin value decreases
significantly in the regions of AR core or with strong currents where
the influence of random errors is suppressed.

Metrics of measuring the force-freeness and divergence-freeness can
be defined in another way by analyzing the residual force in extrapolations~\citep{Jiang2012apj, Malanushenko2014}.
We note that the residual force actually
consists of two parts, the Lorentz force and a force induced by non-zero divergence of the field.
This is because a nonzero $\divB$ can be assumed as being a magnetic
monopole, and analogous to a charge in electric field, it introduces a
force $\vec F=\vec B\divB$ parallel to the field line
\citep{Dellar2001}. Of course this force is un-physical and only results from numerical error.
To define a reference value for these forces, we decompose the Lorentz force,
$(\crlB)\times\vec B$, into two components,
\begin{equation}
  (\crlB)\times\vec B =
  (\vec B\cdot\nabla)\vec B-
  \nabla (B^{2}/2)
\end{equation}
where the two terms on the right hand side are called, respectively, magnetic tension force and magnetic pressure force. These two
components should be balanced in a force-free field, but
each is in general nonzero except in an uniform magnetic
field. Thus a metric of Lorentz force-freeness can be defined by the average
ratio of the Lorentz force to the sum of the magnitudes of the two
component forces, namely,
\begin{equation}
\label{EcrlB}
  E_{\crlB} = \frac{1}{V}\int_{V}\frac{|\vec B \times (\crlB)|}
  {|(\vec B\cdot\nabla)\vec B|+|\nabla(B^{2}/2)|}dV.
\end{equation}
This is identical to Equation (10) of~\citet{Malanushenko2014}, and is
similar to Equation (26) of~\citet{Jiang2012apj} which only used the
magnitude of magnetic-pressure force as denominator.  Similarly, we measure the magnitude of the other force $\vec B\divB$ by
\begin{equation}
  E_{\divB} = \frac{1}{V}\int_{V}\frac{|\mathbf{B}(\divB)|}
  {|(\vec B\cdot\nabla)\vec B|+|\nabla(B^{2}/2)|}dV.
\end{equation}
 These two
metrics $E_{\crlB}$ and $E_{\divB}$ are meaningful in particular if
the extrapolation field is used as initial condition for MHD
simulations~\citep[e.g.,][]{Jiang2013MHD, Kliem2013, Amari2014nat,
  Inoue2014}, because they directly reflect the influence of the
residual force on the numerical MHD system.
Thus we recommend to check these two metrics before using NLFFF
(or any other) solutions to initialize MHD code, and examine carefully the
the related influence. For a typical plasma $\beta \sim 0.01$ in the
lower corona~\citep{Gary2001}, the residual force that can be balanced by the gas pressure should be accordingly $\sim 0.01$ of the magnetic-pressure force, thus the two metrics should be as small as $\sim 0.01$ if the residual force can be considered as negligible.


\begin{table}[htbp]
  \centering
  \caption{Metrics of force-freeness and divergence-freeness
    for the three extrapolation models and the potential field model}
  \begin{tabular}{lllll}
    \hline
    \hline
    Model & CWsin & $\langle |f_{i}|\rangle$  &
    $E_{\crlB}$ & $E_{\divB}$ \\
    \hline
    NLFFF1 & 0.40 & $4.6\times 10^{-4}$ & 0.22 & $4.3\times 10^{-2}$ \\
    NLFFF2 & 0.32 & $3.9\times 10^{-4}$ & 0.16 & $3.6\times 10^{-2}$ \\
    NFFF & 0.80 & $1.2\times 10^{-5}$ & 0.31 & $1.4\times 10^{-4}$ \\
    Potential & 0.86 & $6.0\times 10^{-6}$ & $1.2\times 10^{-4}$ &
    $2.5\times 10^{-4}$ \\
    \hline
  \end{tabular}
  \label{tab:1}
\end{table}

The results for the aforementioned metrics are given in
Table~\ref{tab:1} for all the three extrapolation models as well as a potential field model
that matches the magnetogram. From the results, we
find that all the metrics decrease from NLFFF models 1 to 2, showing
that the NLFFF extrapolations are improved by the preprocessing of the
vector magnetogram, and their values are close to the results in our
previous work for extrapolation of AR~11158 and
AR~11283~\citep{Jiang2013NLFFF}, suggesting that the performance of
CESE--MHD--NLFFF code is not sensitive to different ARs.
The divergence metric $\langle |f_{i}|\rangle$ is on the order of $10^{-4}$, which is consistent with previous results~\citep[e.g.,][]{Metcalf2008, Valori2013, DeRosa2015}.
The value is smaller by one order for the NFFF as it is a superposition of three constant-alpha fields, and it is close to the value for the potential field model.
Comparing $E_{\divB}$ and $E_{\crlB}$ shows that the force induced by the divergence are significantly smaller than the Lorentz
force. We note that the residual force as measured by these two metrics is
larger than what can be neglected if the extrapolated field
is input to an MHD model as a force-free state .
On the other hand, the Lorentz force as measured by both CWsin and $E_{\crlB}$
is much larger in the NFFF
extrapolation than the NLFFFs, as it should be, although the numerical
and measurement uncertainties would still contribute to this value as
described earlier. The divergence-freeness condition is fulfilled much better in the NFFF solution, which is close to the values of that for the potential field model.
This is because the NFFF solution is the sum of three LFFF's for which
the divergence is guaranteed to be as small as the numerical scheme would allow.
Finally we note that CWsin for the potential field is indeed close to 1, which fails to
indicate the force-freeness of a current-free field, as we have discussed,
while the $E_{\crlB}$ gives a reasonable value of $\sim 10^{-4}$, which is as small as the value of $E_{\divB}$.

\begin{figure*}[htbp]
  \centering
  \includegraphics[width=0.9\textwidth]{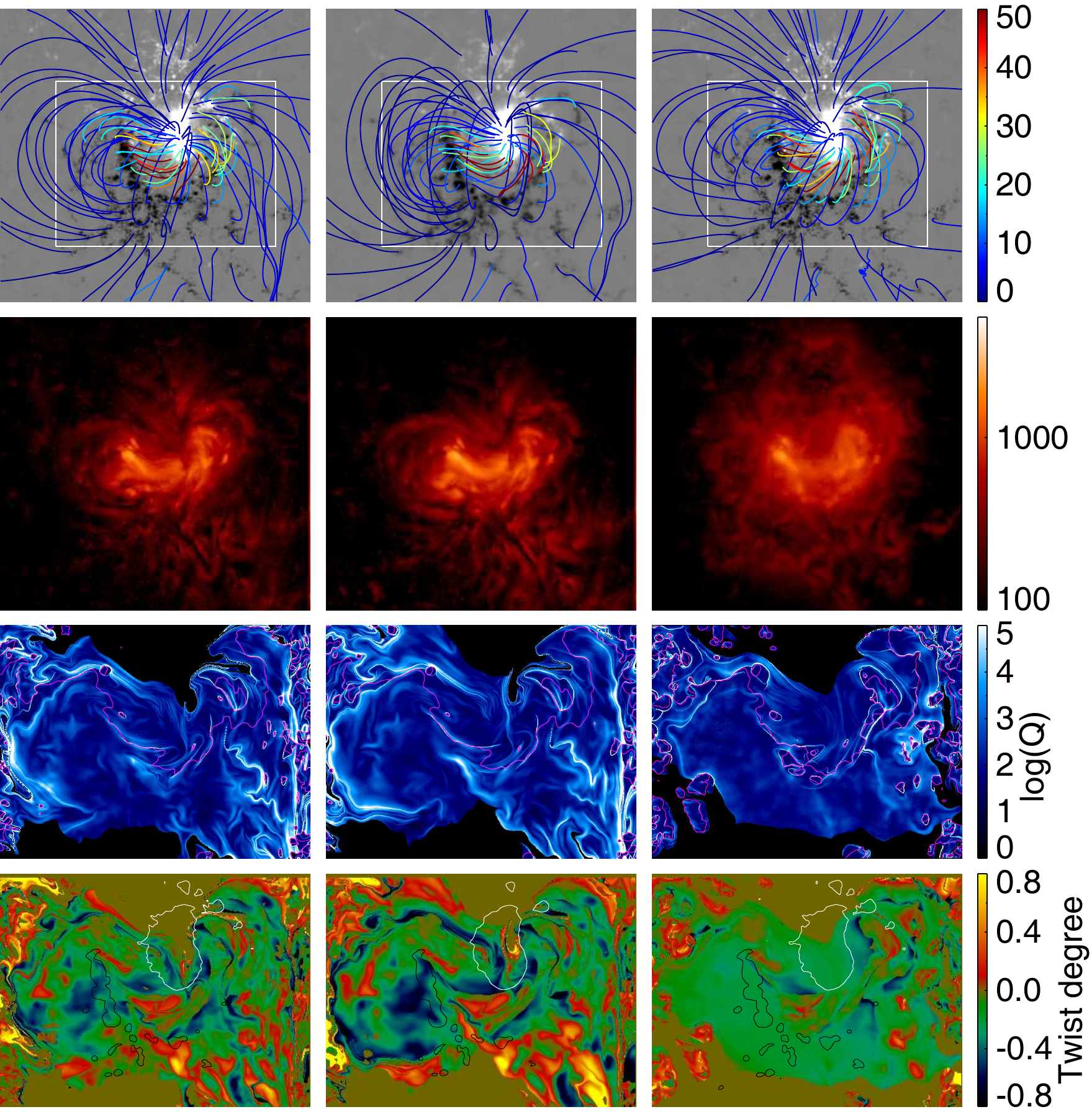}
  \caption{Comparison of the three extrapolation models (columns from
    left to right corresponding to NLFFF1, NLFFF2 and NFFF,
    respectively) by magnetic field lines, electric current
    distributions, magnetic squashing degree and twist degree (rows
    from top to bottom). In the plot of magnetic field lines, each
    field line is color-coded by the mean value of the electric
    current density ($J=|\crlB|$ with unit of G~arcsec$^{-1}$) along
    the line; the backgrounds show the map of $B_{z}$. The current
    distribution is shown by the vertical integration of the current
    density ($\int_{z}J dz$ with unit of G). The distributions of the
    magnetic squashing degree $\log(Q)$ and the
    twist degree are computed at a horizontal slice one pixel above the
    bottom surface of the extrapolation box. Their field of view is
    denoted by the white box in the top panels.  }
  \label{fig:qsl_compare}
\end{figure*}

\subsection{Magnetic field lines, current, and topology}
\label{sec:topo}

In \Fig~\ref{fig:qsl_compare} we compare the magnetic lines, electric current distributions, and magnetic topology for three magnetic fields. The columns
from left to right are respectively for NLFFF1, NLFFF2, and NFFF. The
magnetic field lines are sampled and shown in the first row of the
figure. Each field line is color-coded by the mean value of the
electric current density ($J=|\crlB|$ in G~arcsec$^{-1}$)
along the line. The second row of the figure shows a vertical
integration of the current density ($\int_{z}J dz$ with unit of
G). Overall an inverse-S shape of strong current can be seen, especially in
the NLFFFs, which is very roughly consistent with the AIA 94~{\AA} image
(\Fig~\ref{fig:fulldisk}c). The strongest current are associated with the
magnetic field lines in the core region, which corresponds to the
brightest loops in AIA 94~{\AA} image. The magnetic field lines
of all the models here are mostly sheared arcades rather than fully S
shapes, and show no presence of the strongly twisted magnetic flux
rope that is found in the extrapolation by~\citet{ZhaoJ2016}. Our
results seem to agree with the observations of AIA, from which the
loop tracing code identifies approximately formed S-shaped loops.
From a visual inspection of the magnetic field lines of the models, it appears that
in the core region the NLFFFs possess stronger shear than the NFFF, contrarily, in
the enveloping field region, the latter has slightly stronger
sheared field lines than the former ones. Consistently, the current
distribution of the NFFF is more diffused than that of the NLFFF
models.

To further compare the magnetic topology, we compute the magnetic
squashing degree at the bottom surface. The squashing degree, or $Q$~factor, is a quantity measuring shape distortions of elemental flux tubes
based on the field-line mapping~\citep{Demoulin1996,Titov2002}.  Specifically, starting at footpoint $(x,y)$, the other footpoint of a closed magnetic field line is denoted by $( X(x,y), Y(x,y) )$ for a given coronal field model. Then $Q(x,y)$ is given by
\begin{equation}
  \label{eq:Q}
  Q = \frac{a^{2}+b^{2}+c^{2}+d^{2}}{|ad-bc|}
\end{equation}
where
\begin{equation}
  a = \frac{\partial X}{\partial x},\ \
  b = \frac{\partial X}{\partial y},\ \
  c = \frac{\partial Y}{\partial x},\ \
  d = \frac{\partial Y}{\partial y}.
\end{equation}
Thus a small value of $Q$ indicates that a infinitesimal circle at one foot-point
is mapped to a circle at the other foot-point, while a
very high value (e.g., $>100$) indicates extreme distortion of the circle, which indicates
steep gradient of the field-line mapping
 that occurs in magnetic
quasi-separatrix layers (QSLs). Thus the $Q$ factor is useful for searching for
important topological structures like separatrices and QSLs.
They prove to be relevant to the studies of reconnection sites in the corona and
thus physics of flares. For computing $Q$, we use the approach recently proposed by~\citet{Pariat2012}, which is computationally efficient and can be used to
compute $Q$ inside the 3D domain. The
third row of \Fig~\ref{fig:qsl_compare} shows the maps of $Q$~factor. Comparing  these three models, we find limited difference
between them, and there appears to be no well-shaped QSL defining the
boundary of the shear core and the enveloping flux.


In the last row of \Fig~\ref{fig:qsl_compare} we show the map of twist
 ($T_n$) of magnetic field lines~\citep[e.g.,][]{Inoue2011,
  LiuR2016}, which is defined by
\begin{equation}\label{Tn}
  T_n = \frac{1}{4\pi}\int_L \frac{(\crlB)\cdot\vec B}{B^2} dl.
\end{equation}
where the integral is taken along each closed field line. The magnetic
twist $T_n$ measures number of turns two infinitesimally
close field lines make about each other~\citep{Berger2006}. Both signs
of twist are seen in the whole region, and the majority of the
core field has a negative value of twist, i.e., a left-handed
twisting, and with a relatively small value $T_n \leq 1$. This is
consistent with the absence of strongly-twisted flux rope in the
models.
Comparison of the two NLFFF models shows
that the preprocessing also results in clear increase of the
magnetic twist. The non-force-free model gives a much more
weakly twisted field than the NLFFF models, and the distribution of
twist degree appears much more even.
A further study of
the magnetic topology and distribution of twist degree and their
relation with the eruption is left to a future paper.

\begin{figure*}[htbp]
  \centering
  \includegraphics[width=0.9\textwidth]{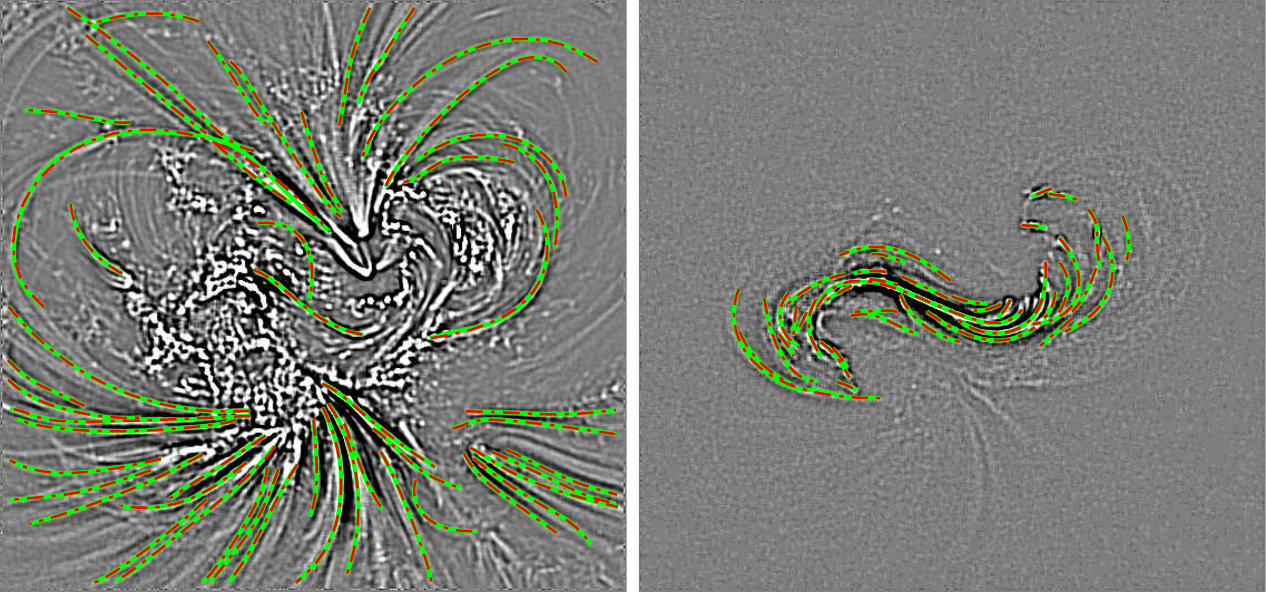}
  \caption{
   Fitting of the OCCULT-2 traced loops (thick green curves)
   by 2D B{\'e}zier splines (dashed red curves). The background are
   bandpass filtered AIA images, left for AIA 171~{\AA} and right for AIA 94~{\AA}.
   }
  \label{fig:bezier_fitting}
\end{figure*}

\begin{figure*}[htbp]
  \centering
  \includegraphics[width=0.9\textwidth]{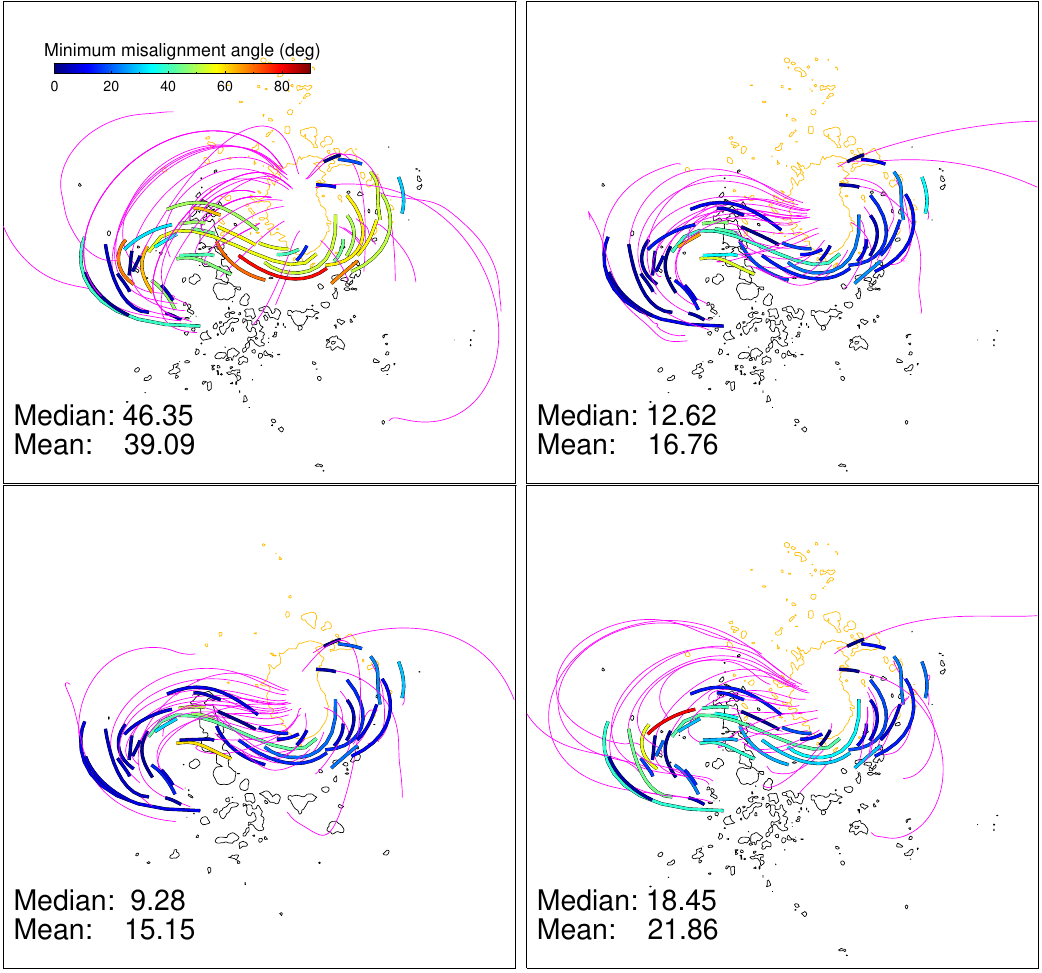}
  \caption{Minimum misalignment angles between the magnetic field
    model and the B{\'e}zier splines (for AIA 94~{\AA} loops) when
    extended to 3D. The four panels clockwise from the top left to the
    bottom left are results, respectively, for the potential magnetic
    field, the NLFFF1, the NFFF, and the NLFFF2
    models. Each loop, as shown by the thick curves, is color-coded by
    the value of its misalignment angle ($0-90^{\circ}$) as indicated
    by the colorbar in the first panel. Also a magnetic field line
    (shown by thin curves) is traced from the mid-point of each
    loop. The contour lines are plotted for $B_{z}$ at $-500$~G
    (colored as black) and $500$~G (yellow). For each model, the
    median and mean values of the misalignment angles for all the
    loops are shown.}
  \label{fig:AIA94_fitting}
\end{figure*}

\begin{figure*}[htbp]
  \centering
  \includegraphics[width=0.9\textwidth]{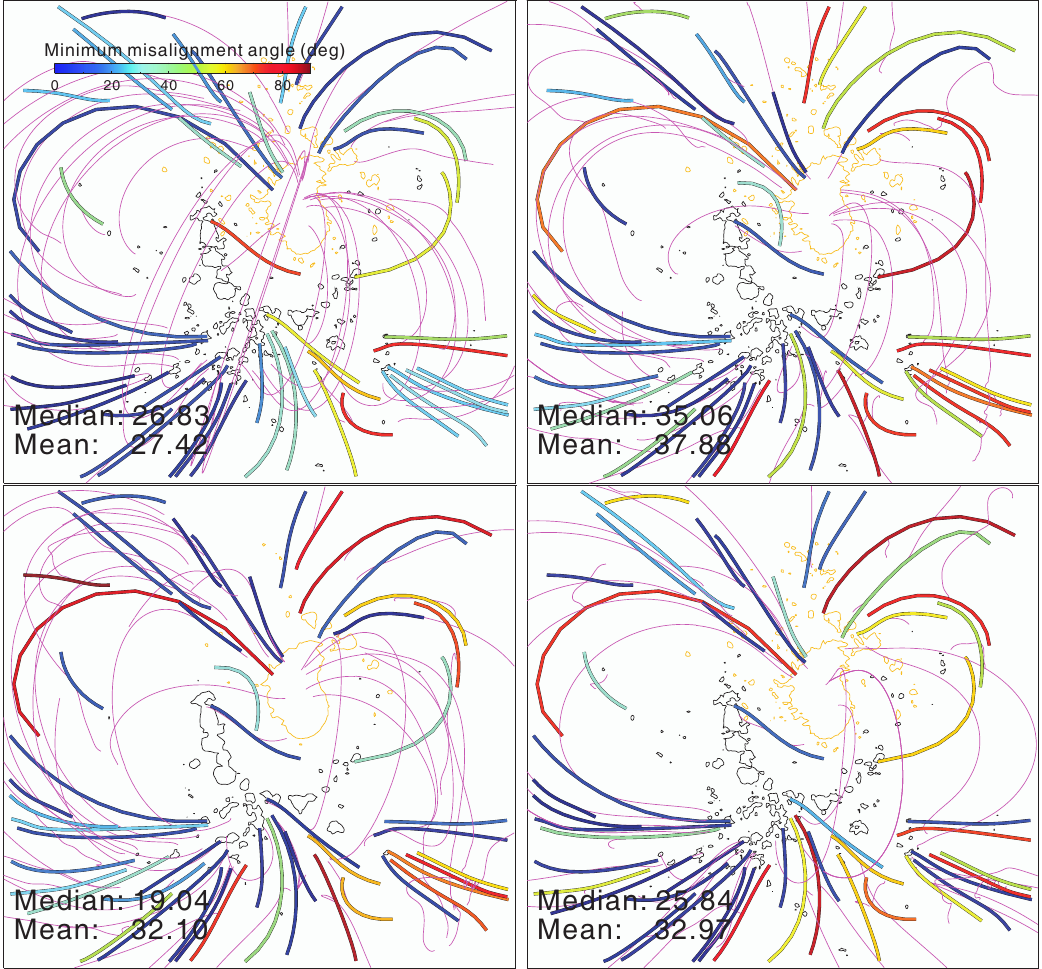}
  \caption{Same as \Fig~\ref{fig:AIA94_fitting} but for AIA 171~{\AA}
    loops.}
  \label{fig:AIA171_fitting}
\end{figure*}

\subsection{Misalignment of the magnetic field lines and AIA loops}
\label{sec:misalignment}

Now we report the results of loop fitting to the magnetic field models
as described in Section~\ref{sec:loop}. Here we apply the procedures
of~\citet{Gary2014} to two AIA channels, i.e., 94~{\AA} and 171~{\AA},
which represent two separate features of the AR field. The
94~{\AA} shows loops at the AR core that are presumably associated with
significantly non-potential magnetic field, and they are usually compact and
short, while the 171~{\AA} loops correspond to the enveloping field that
is often close to a potential state.  As
described in Section~\ref{sec:loop}, the OCCULT-2 traced loops from
the AIA images are first fitted by 2D 4-point B{\'e}zier splines.  The
results are shown in \Fig~\ref{fig:bezier_fitting}. As can be seen,
although the 4-point B{\'e}zier fitting is previously only applied to
AIA 171~{\AA} loops, here it works also pretty well for the AIA
94~{\AA} loops, some of which exhibit slightly S shapes. Then, using the method
described in Section~\ref{sec:loop}, the
B{\'e}zier splines (or loops) are extended to 3D by minimizing the
misalignment angle between the loops and the 3D magnetic
field. \Figs~\ref{fig:AIA94_fitting} and \ref{fig:AIA171_fitting} give
the results of such a minimization of the  misalignment angle for each loop, and the
different panels in the figures are for different extrapolation models
of magnetic field. For convenience
of comparison, the loops are shown by the thick curves color-coded by
value of corresponding misalignment angle with the magnetic field, and
for each loop, a magnetic field line is traced passing through the
mid-point of the loop, which are shown by the thin curves. The median
and mean values of all the misalignment angles for each magnetic field
model are also shown in the figures.

We find that model NLFFF2 gives the best results in fitting to
the AIA~94~{\AA} loops as it has the smallest values for both the
median and mean of the misalignment angles. For the AIA~171~{\AA}
loops, NLFFF2 also gives the smallest median value, while the
smallest average value is given by the potential field model.
Comparing the results for the two AIA channels, we notice that the fitting
of the modeled field lines to loops is much better for the 94~{\AA}
than for 171~{\AA}. In particular in the best model (NLFFF2), the
misalignment angles for the 94~{\AA} loops, which have median value of
$\sim 10^{\circ}$ and mean value of $\sim 15^{\circ}$, are less than half of those for the 171~{\AA} loops
($\sim 30^{\circ}$). For 94~{\AA} loops, all the
non-potential models give significantly smaller value of misalignment
than the potential field model, which confirms that these loops are
associated with magnetic field of significant
non-potentiality. Improvement from preprocessing of the magnetogram
for NLFFF extrapolation is also confirmed by the decrease of the
misalignment angles.

It is worthy noting that our best results for the AIA 94~{\AA}
loops are on the order of the results
using coronal loop forward-fitting NLFFF code shown
in~\citet{Aschwanden2013ApJ}, who found that the 3D misalignment angle
amounts to an average value of $19^{\circ}\pm 3^{\circ}$ for their
studied ARs.  However, for the 171~{\AA} loops, our extrapolation
models perform, at the best, only close to the potential field model
(with misalignment of $\sim 30^{\circ}$).  Such value is comparable to
the misalignment angles ($24^{\circ}-44^{\circ}$) as resulted from a
comparison of various NLFFF models for AR~10953 by~\citet{DeRosa2009},
who also found that those NLFFF models perform no better than
potential model. The NLFFF1 without preprocessing gives even larger
misalignment angle than the potential model, and it is slightly improved from
the preprocessing. The different values for
94~{\AA} and 171~{\AA} show that the CESE--MHD--NLFFF code can
reconstruct much more reliably the AR core field (which is
significantly non-potential) than the envelope field (which is
presumably close to potential). The large misalignment of the
extrapolated field with the 171~{\AA} loops is because the field
lines corresponding to the long loops do not relax sufficiently during
the MHD-relaxation process. Since in the code the relaxation speed is
uniform and the relaxation time is thus proportional to the field line
length, the long field lines need more time to relax than the short
ones. Moreover, the long field lines often extend close to or reach
side and top boundaries of the computational volume and are subject to the
boundary effects, as all these boundaries are fixed with the initial potential field (however note that the computational volume is larger than the field-of-view of the magnetogram).
Another factor which might be partially responsible for this effect is the departure from the
Cartesian coordinates assumption further away from the core. Thus a further optimization of the relaxation
process with realization of spherical geometry is needed for a better reconstruction of the enveloping field.


\begin{table}[htbp]
  \centering
  \caption{Magnetic flux, energy contents and relative helicity of the
    extrapolation models. Units are respectively, $10^{22}$~Mx for magnetic flux,
    $10^{32}$~erg for magnetic energy, and $10^{43}$~Mx$^{2}$ for
    magnetic helicity.}
  \begin{tabular}{llllllll}
    \hline
    \hline
    Model & $\Phi$ & $E_{\rm tot}$ & $E_{\rm pot}$ & $E_{\rm free}$ &
    $E_{\rm free}/E_{\rm pot}$ & $H$ & $H/\Phi^{2}$ \\
    \hline
    NLFFF1 & 3.41 & 11.5 & 11.7 & -0.18 & -1.5\% & -1.60 & $-1.38\times 10^{-2}$ \\
    NLFFF2 & 3.18 & 11.1 & 10.0 & 1.10  & 10.9\% & -2.03 & $-2.00\times 10^{-2}$ \\
    NFFF & 3.41 & 14.8 & 11.8 & 3.02  & 25.6\% & -2.36 & $-2.04\times 10^{-2}$ \\
    \hline
  \end{tabular}
  \label{tab:2}
\end{table}

%
\subsection{Magnetic energy and helicity}
\label{sec:erg}

In Table~\ref{tab:2} we list the magnetic energy contents and relative
helicity for the extrapolation solutions. Both the total magnetic
energy $E_{\rm tot}$ and the corresponding reference potential-field
energy $E_{\rm pot}$ are computed, from which the free magnetic energy
$E_{\rm free}=E_{\rm tot}-E_{\rm pot}$ can be derived.  The reference
potential fields are computed numerically by solving the Laplace
equation for the potential with Neumann boundary conditions based on
the normal component of $\vec B$ on all six boundaries of the analysis
volume. As a consequence of each solution field having its own
boundary value, there are separate reference potential fields for each
of the solution fields, and thus the potential-field energy contents
are different among the models. The relative magnetic helicity $H$ is
defined following~\citet{Berger1984}
\begin{equation}
  \label{eq:H}
  H = \int_{V} (\vec A+\vec A_{\rm pot})(\vec B-\vec B_{\rm pot})dV
\end{equation}
where $\vec A$ is vector potential of the magnetic field, i.e., $\vec
B = \nabla\times \vec A$ and $\vec B_{\rm pot}$ is the reference
potential field and also $\vec B_{\rm pot} = \nabla\times \vec A_{\rm
  pot}$. The calculation of $H$ is made using a rapid method
developed by~\citet{Valori2012}. For reference, we also list values of
the total unsigned magnetic flux $\Phi$ in the table.

Clearly the free energy increases from NLFFF1 to NLFFF2. Such an
increase of non-potentiality of the solutions is due to the
improvement of the NLFFF extrapolations. Here the free
energy for NLFFF1 is even un-physically negative. We note that energy of the extrapolated field being lower than the potential field energy has been reported previously, for instance, some solutions given in Table~3 of \citet{Metcalf2008} and Table~1 of \citet{Schrijver2008a}.
\citet{Valori2013} shows that such problematic extrapolation actually
reflects the violation of Thomson's theorem (that the total energy of a system can be expressed as the sum of its potential energy and its free energy) due to the finite divergence of the extrapolation field rather than the non-zero Lorentz force. The finite divergence is unavoidably induced in the course of seeking a force-free field with an inconsistent boundary condition (i.e., the un-preprocessed photospheric field).
Since the MHD-relaxation code attempts to construct force-free solution by reducing the Lorentz forces in the computing volume, while the boundary condition is incompatible with the force-free equation, the reduction of the Lorentz forces is at the expense of the solenoidal condition. In such case, the more inconsistent the boundary is, the higher  the divergence of the solution will be, and Thomson's theorem will be more severely violated, which could result in a total energy lower than the potential energy. The inconsistency of the photospheric field can be partly reduced by preprocessing.
After preprocessing, the code outputs free
energy on the order of $10^{32}$~erg, which could
energize a major flare. Interestingly, the NFFF model outputs a
significantly larger free energy content, which is about three times
of that from NLFFF2, This might be due to the fact that NFFF is a combination
of LFFFs whose current is distributed more evenly and result in a field
much more non-potential than the NLFFFs.


All the models have negative helicity, which complies with the hemispheric
chirality-rule~\citep{Seehafer1990, Pevtsov1995}, i.e., active regions in
southern and northern hemispheres tend to have negative and positive
helicities respectively. A negative
helicity indicates a left-handed twisting of the magnetic field lines,
and can be easily confirmed by inspecting the magnetic field lines
and their directions (see \Fig~\ref{fig:qsl_compare}) and also in the
twist degree map~(\Fig~\ref{fig:qsl_compare}), which shows that the  majority
of the values is negative. Consistent with the increasing of free energy
from NLFFF1, NLFFF2 to NFFF, the relative helicity contents also
increase.

\section{Discussion}
\label{sec:con}

Due to the absence of direct measurements, to determine the  coronal magnetic field one has to extrapolate the photospheric field. Critically assessing and comparing the results for
different extrapolation codes is an important task for developing a
reliable coronal magnetic field model consistent with observations and physics. In this
paper, we have made a comprehensive assessment and comparison of an
NLFFF code (CESE--MHD--NLFFF) and a non-force-free code. For this, we
extrapolated the coronal magnetic field for a sigmoidal AR near its
central meridian passage immediately before a major flare. Vector
magnetograms of the HMI SHARP dataset are used, and three field models
are calculated including the CESE--MHD--NLFFF extrapolations using the
magnetogram with and without preprocessing and the NFFF
extrapolation. The extrapolation solutions are evaluated and compared
in different ways including the residual Lorentz force and divergence,
the magnetic topology and energy/helicity contents. In particular, we
use a method recently proposed by \citet{Gary2014} to compute the
misalignment angle of the model field lines with the 3D traces of coronal loops identified from two AIA images (wave lengths of
$94$~{\AA} and 171~{\AA}), respectively, which previously have not been tested for these codes.

The extrapolations show that the magnetic field lines in the AR's core
is consistent with sheared arcades forming a sigmoidal shape in agreement
with the observed AIA~94~{\AA} loops. Twist of the field lines are
mostly below one full turn, and also there seems to be no QSL that
marks a clear presence of magnetic flux rope. It is found that the
best extrapolated field matches the AIA~94~{\AA} loops with misalignment angles of mean value $\sim 15^{\circ}$ and median value $9^{\circ}$, which are much smaller than that for the 171~{\AA} loops (mean value of $\sim~32^{\circ}$ and median value of $\sim~19^{\circ}$). Interestingly, the misalignment angles for AIA~94~{\AA} loops are even
comparable with those by coronal-loop forward-fitting
method~\citep{Aschwanden2013ApJ}, suggesting that the CESE--MHD--NLFFF
can reproduce the magnetic configuration at the core
region of an AR.
On the other hand, the
171~{\AA} loops are not well reproduced, which means that an
improvement of the code is necessary for further relaxation of
the long field lines.
By comparing the NLFFF extrapolations made using both preprocessed and not preprocessed input data, we confirm the preprocessing of vector
magnetogram improves considerably the extrapolation result.  For the
best results, the residual Lorentz force is as large as $15\%$ in normalized units (see above) while the residual divergence is one order of magnitude smaller. When comparing the NLFFFs
and the NFFF results, it is found that the currents of the
field by the NFFF is distributed more evenly than in the NLFFFs for this particular AR. In the future we plan to extend the comparison study with coronal loops to a number of AR samples, as well as including the magnetic field
results from a data-driven MHD model~\citep{Jiang2016NC}. Experiments by using NFFF solutions as initial conditions to MHD simulations are also underway.


\acknowledgments

This work is jointly supported by National Natural Science Foundation of China
(41604140), China Postdoctoral Science Foundation funded project (119103S277), and
the Specialized Research Fund for State Key Laboratories.
C.J. acknowledges support by
 National Natural Science Foundation of China
(41531073, 41374176, 41574170, 41231068, and 41574171).
Data from observations are courtesy of NASA {SDO}/AIA and
the HMI science teams. We thank the anonymous referee for helpful comments on the
manuscript. We were all saddened by the sudden passing of Dr.~S.~T.~Wu. We dedicate this work to ST in memory of his life-long achievement, his mentorship, and his vision for numerical modeling of solar coronal magnetic field.


\end{CJK*}
\end{document}